\documentclass[aps, a4paper,twocolumn]{revtex4-2}
\usepackage[paperwidth=210mm,paperheight=297mm,centering,hmargin=1.5cm,vmargin=2cm]{geometry}

\usepackage[english]{babel}
\usepackage{graphicx}
\usepackage{xcolor}
\usepackage{physics}
\usepackage{hyperref}
\usepackage{amssymb}
\usepackage{amsmath}
\usepackage{tabularx}
\usepackage{multirow}

\definecolor{naturered}{RGB}{228,28,37}
\definecolor{natureorange}{RGB}{241,127,26}
\definecolor{natureyellow}{RGB}{255,203,0}
\definecolor{naturegreen}{RGB}{146,205,28}
\definecolor{naturemidblue}{RGB}{25,141,241}
\definecolor{natureblue}{RGB}{1,50,255}

\newcommand{\GHZ}[1]{$\mathrm{GHZ}_{#1}$}
\def\defeq{\mathrel{\mathop:}=} 

\begin{document}

\title{Anonymous and secret communication in quantum networks} 

 \author{Christopher Thalacker$^{1,2}$, Frederik Hahn$^{3}$, Jarn de Jong$^{4}$, Anna Pappa$^{4}$, Stefanie Barz$^{1,2}$}
 \affiliation{%
$^1$Institute for Functional Matter and Quantum Technologies, Universit{\"a}t Stuttgart, 70569 Stuttgart, Germany \\
$^2$Center for Integrated Quantum Science and Technology (IQST), Universit{\"a}t Stuttgart, 70569 Stuttgart, Germany\\
$^3$Dahlem Center for Complex Quantum Systems, Freie Universit{\"a}t Berlin, 14195 Berlin, Germany\\
$^4$Electrical Engineering and Computer Science Department, Technische Universit{\"a}t Berlin, 10587 Berlin, Germany}%

\begin{abstract}
Secure communication is one of the key applications of quantum networks.
In recent years, following the demands for identity protection in classical communication protocols, the need for anonymity has also emerged for quantum networks.
Here, we demonstrate that quantum physics allows parties---besides communicating \textit{securely} over a network--- to also keep their \textit{identities} secret.
We implement such an anonymous quantum conference key agreement by sharing multipartite entangled states in a quantum network.
We demonstrate the protocol with four parties and establish keys in subsets of the network---different combinations of two and three parties---whilst keeping the participating parties anonymous.
We additionally show that the protocol is verifiable and run multiple key generation and verification routines.
Our work thus addresses one of the key challenges of networked communication: keeping the identities of the communicating parties private.

\end{abstract}

\maketitle

Quantum communication has developed from first proof-of-principle demonstrations to real-world applications. 
Since the first proposals of using quantum physics to establish secret keys between two parties more than three decades ago~\cite{Bennett1984,Ekert1991,Bennett1992}, numerous demonstrations of quantum key distribution have been performed. Starting from proof-of-concept experiments, intermediate and large-scale quantum networks spanning thousands of kilometres have recently been developed~\cite{Ursin2007, Peng2007,Peev2009,Sasaki2011,Liao2017, Pugh2017, Chen2021}. These experiments use single photons or entangled photon pairs to exchange a secret key between two parties and are known as \textit{quantum key distribution} (QKD) protocols~\cite{Krenn2016, Flamini2018}.

With growing complexity of quantum networks, new possibilities arise for multiparty protocols that go \textit{beyond} bipartite secure communication. 
One example is \textit{conference key agreement} (CKA), a cryptographic primitive that allows parties in a large-scale network to jointly establish a shared secret key\,\cite{Murta2020}. 
Several approaches have been proposed with different requirements for quantum resources, ranging from bipartite to \textit{multipartite} quantum states shared over the network. Interestingly, sharing multipartite states was shown to be more efficient in specific cases, e.g.~for networks that have bottlenecks~\cite{Epping2017}. Recently, experimental implementations of CKA protocols have been performed~\cite{Proietti2020}.

\begin{figure}
         	\includegraphics[width=0.44\textwidth]{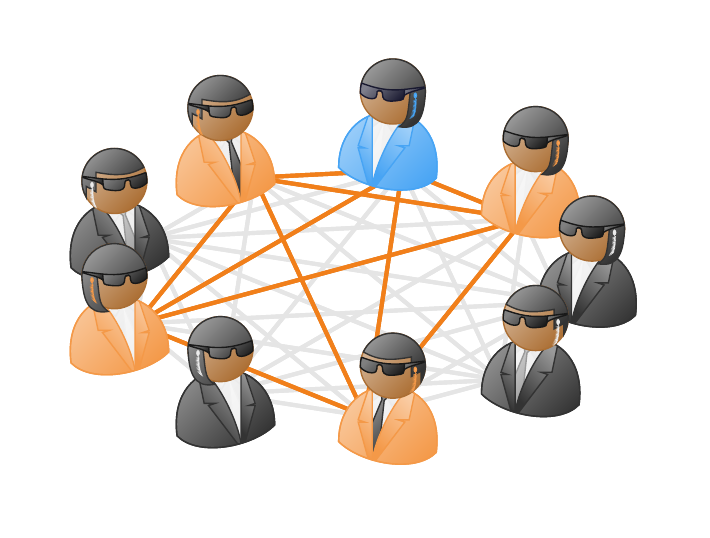}
         	\caption{\textbf{Anonymous Conference Key Agreement.} A sender (\textcolor{naturemidblue}{\textbf{blue}}) aims to establish a secret key with $m$ participants (\textcolor{natureorange}{\textbf{orange}}) while their identities remain secret, both from each other and from the remaining non-participants (\textcolor{black}{\textbf{black}}). 
         	}
         \label{fig:ACKA}
	
\end{figure}

Beyond the security of key generation, quantum networks open up a wide range of possibilities regarding other aspects of secure communication. One such aspect is \emph{anonymity}~\cite{Hahn2020, Unnikrishnan2019, Murta2020, Proietti2020, Huang2020}. 
In a broad range of cases, internet users put much effort into keeping their activities and identities secret. As classical networks are replaced by their quantum counterparts, anonymity will likewise be a vital requirement for networked quantum communication.

Various levels of anonymity can be envisioned, and there exist multiple real-life scenarios 
where the identities of one or more of the communicating parties need to be kept secret.
One example is whistle blowing, where it is imperative that the identity of the sender is kept secret. 
Combining anonymity with the requirement for private communication between multiple parties leads to \textit{anonymous conference key agreement} (ACKA)~\cite{Hahn2020}. Here, the goal is to establish a secret key between several parties across a larger network in such a way that their identities are hidden from everyone else in the network, including each other, and are only known to the initiating party~(see~Fig.~\ref{fig:ACKA}).

In this work, we demonstrate how to anonymously establish a secret key in a quantum network of four parties using multipartite entanglement.
We focus on a protocol first proposed in~\cite{Hahn2020}, extending the work of~\cite{Unnikrishnan2019, Yang2020} on bipartite key generation.

Our protocol works as follows: first, each party is notified whether they are meant to participate in the key exchange or not~\cite{Broadbent2007}. 
Then, Greenberger-Horne-Zeilinger (\GHZ{}) states are repeatedly shared with all parties in the network, such that each party receives one qubit from every shared state~\cite{Greenberger1989, McCutcheon2016, Wang2016, Wang2018, Zhong2018}. 
A few of these shared states are used to generate a secret key; the majority is used to detect an eavesdropper or any deviation of the parties from the protocol, making the protocol also \textit{verifiable}.
In our implementation, we demonstrate six different configurations for anonymously establishing a secret key between a sender and one or two participants in a four-partite network. 

%
\begin{figure*}[tb]
\includegraphics[width=1.0\textwidth]{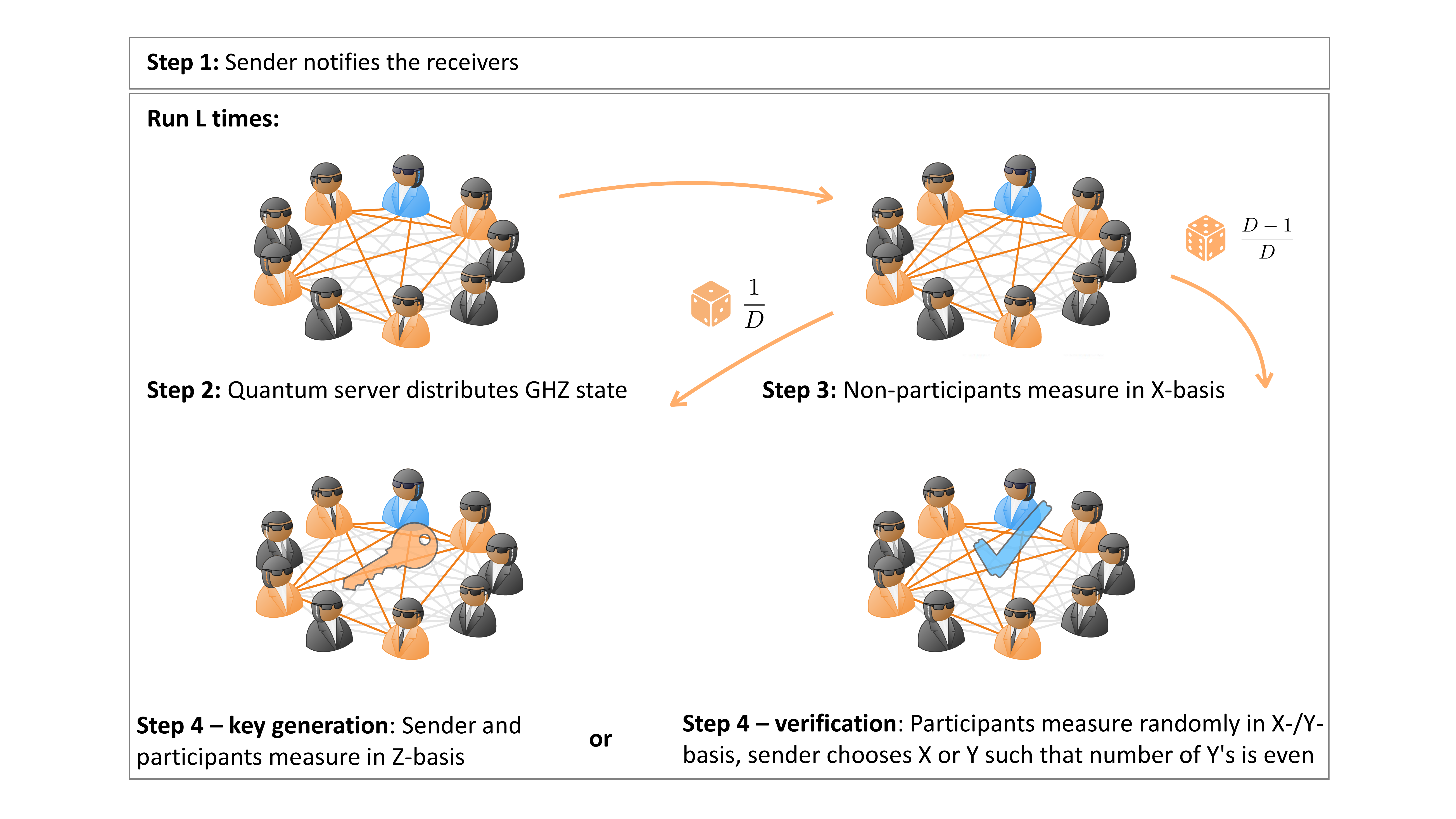}
	\caption{\textbf{Sketch of the anonymous conference key agreement protocol.} A sender (\textcolor{naturemidblue}{\textbf{blue}}) notifies $m$ participants (\textcolor{natureorange}{\textbf{orange}}) and establishes a secret key while their identities remain secret, both from each other and from the remaining non-participants (\textcolor{black}{\textbf{black}}). See main text for detailed explanation of the protocol.}
	\label{fig:flowchart}
\end{figure*}
%

%
\section{The protocol for anonymous conference key agreement}
%
%

The goal of the protocol is for a sender to anonymously establish a key with $m$ participants of their choice while they form part of a larger network of $n$ parties. The network is able to distribute a state
\begin{align}\label{eqn:ghztheory}
\ket{\mbox{GHZ}_n}\defeq\tfrac{1}{\sqrt{2}}\left(\ket{0}^{\otimes n}+\ket{1}^{\otimes n}\right)
\end{align}
between all $n$ parties (cf.~Fig.~\ref{fig:flowchart}).

The protocol starts with the $n$ parties running a classical notification protocol that allows the sender to anonymously notify $m$ participants---and thereby implicitly the non-participants~\cite{Broadbent2007}. This requires private classical communication between all pairs of parties. Once the participants are notified, the remainder of the protocol is repeated $L$ times. In each round, a \GHZ{n} state is distributed to the $n$ parties and the non-participants measure their qubits in the $X$-basis. This results in a \GHZ{m+1} state between only the sender and the participants, with an additional phase of $\pm 1$ depending on the parity $\Delta$ of the measurement outcomes of the non-participants, i.e.
\begin{align}\label{eqn:ghztheorymplus1}
\ket{\mbox{GHZ}_{m+1, \Delta}}=\tfrac{1}{\sqrt{2}}\left(\ket{0}^{\otimes {(m+1)}}+(-1)^{\Delta}\ket{1}^{\otimes {(m+1)}}\right).
\end{align}

In order for this phase to be compensated, the non-participants  publicly announce their measurement outcomes---while the participants hide their identity by announcing random bits. The sender can tell everyone apart and can now apply a local correction to her qubit, thereby canceling this phase. 
This allows the sender and the participants to anonymously `extract' a smaller \GHZ{m+1} state shared only between themselves. This state can consequently be used to anonymously generate a shared secret key due to its inherent correlations when measured in the computational basis.

Note that, however, the non-participants could have diverged from their expected behaviour---for instance by not measuring their qubits and announcing instead a random bit. This would allow them to be part of the key-generating network, without being noticed. To prevent this, the participants use a large percentage of the shared \GHZ{m+1} states for anonymous verification. 
In these \texttt{Verification} rounds, the participants perform measurements of random stabilisers on the \GHZ{m+1} state, whose outcome parity is known to be one. All participants announce their measurement bases and outcomes, while the non-participants announce two random bits concealing their roles. Only the sender is able to distinguish between these announcements and can thus validate the results. 
By performing $D-1$ randomly chosen \texttt{Verification} rounds out of every $D$ rounds in total, the sender can thus estimate the closeness of the distilled state to the ideal \GHZ{m+1} state~\cite{Pappa2012}. This guarantees that the resulting key is provably secure and secret, while also preserving anonymity as no communication takes place during the \texttt{KeyGen} rounds.  
We obtain a key of length $L/D$ which is provably secure and the participants remain anonymous. Which of the rounds will be \texttt{Verification} rounds and which \texttt{KeyGen} rounds is decided by a $1/D$-biased public source of randomness.

%
\section{Experimental Implementation}
%
%

We demonstrate anonymous verifiable conference key agreement in a network with four parties. 
We generate four-photon \GHZ{4} states encoded in polarisation (H = 0, V = 1) using an all-optical setup as shown in Fig.~\ref{figure3}~(a) and in more detail in Fig.~\ref{figure5} in the Appendix. The generated states have a fidelity of $F=0.85(\pm0.02)$, estimated via quantum state tomography~\cite{James2001}.

The four-photon state allows us to demonstrate various configurations of the anonymous conference key agreement: a sender and two receivers, and a bipartite protocol with a sender and one receiver. 
In all configurations, as shown in Fig.~\ref{figure3}~(b), all participating parties remain anonymous.
As a consequence, the non-participants do not know which and how many parties are in the end sharing a secret key, since they cannot distinguish between the different configurations.

We assume that the classical \texttt{Notification} protocol has already been performed and we start by distributing the \GHZ{4} states to all parties. 
First, the non-participants perform a measurement in the $X$-basis, resulting in a smaller \GHZ{3} or \GHZ{2} state.

\begin{figure}[t]
   	\includegraphics[width= 0.48\textwidth]{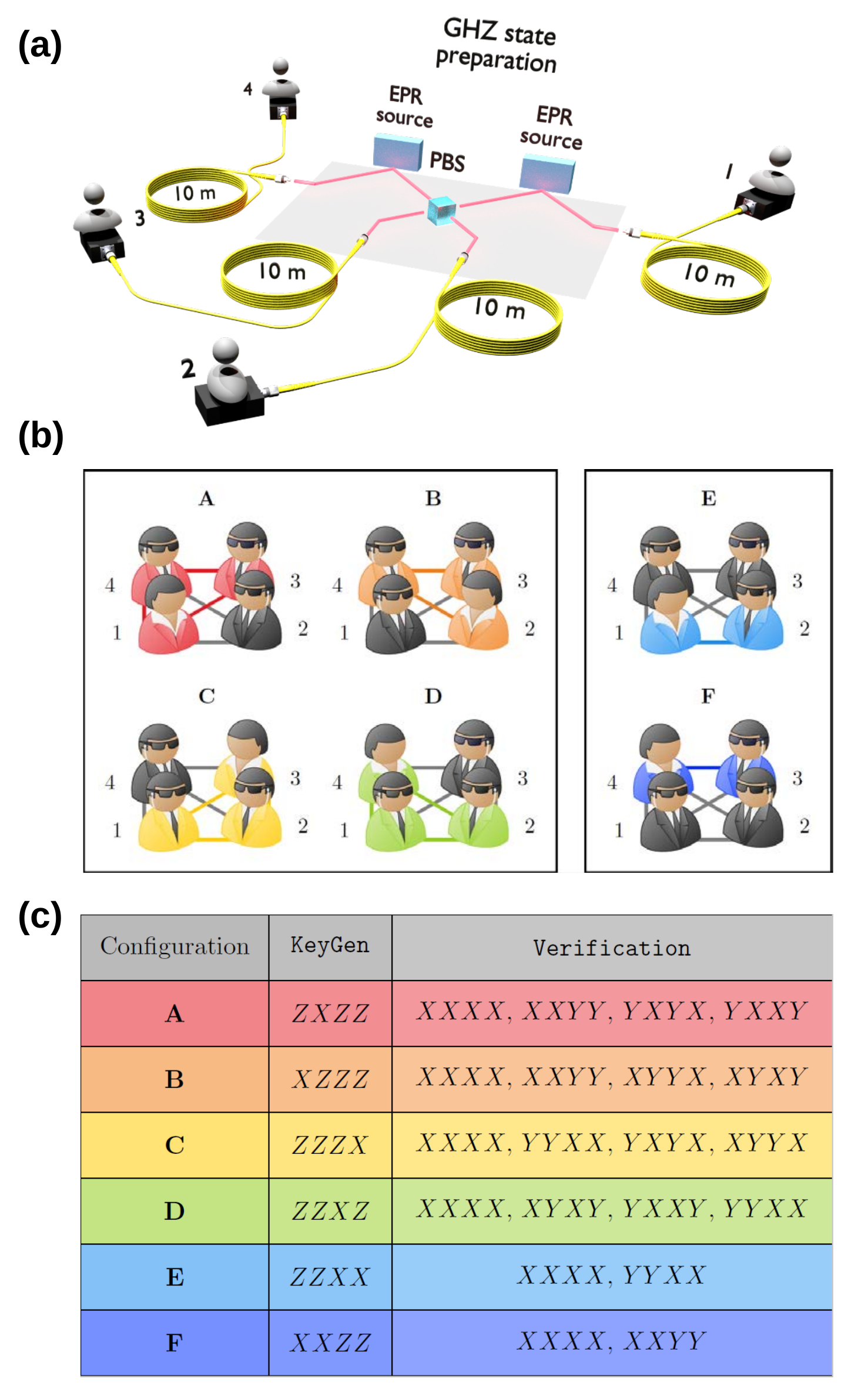}
         \caption{The setting of our implementation. (\textbf{a)} Sketch of the experimental setup with four receivers. Two polarisation-encoded EPR photon pairs are generated and two photons---one of each pair---are superimposed on a PBS to generate a four-photon \GHZ{4} state. This state is then distributed to the four parties. \textbf{(b)} Different network configurations implemented in our demonstration. The sender and the participants are highlighted in colour, while the non-participants are displayed in black.
				\textbf{(c)} Resulting measurement settings for each configuration and type of round, \texttt{KeyGen} or \texttt{Verification}.}
		\label{figure3}
\end{figure}

We then randomly choose between \texttt{KeyGen} rounds and \texttt{Verification} rounds.
In a \texttt{KeyGen} round, all participants measure in the $Z$-basis and exploit the correlations of the \GHZ{} state for establishing a shared key.
In the \texttt{Verification} rounds, each participant randomly measures in the $X$- and $Y$-basis and announces their basis together with their measurement outcome, while the sender announces random bits. Then, the sender performs either an $X$- or $Y$-measurement so that the total number of $Y$-measurements is even. 

Figure~\ref{figure3} (c) shows the different measurement configurations of each setting according to the protocol.
In our implementation, all four measurements are implemented at the same time using motorised half-wave and quarter-wave plates, together with a polarising beam splitter. \\

\begin{figure*}
        	\includegraphics[width=1.0\textwidth]{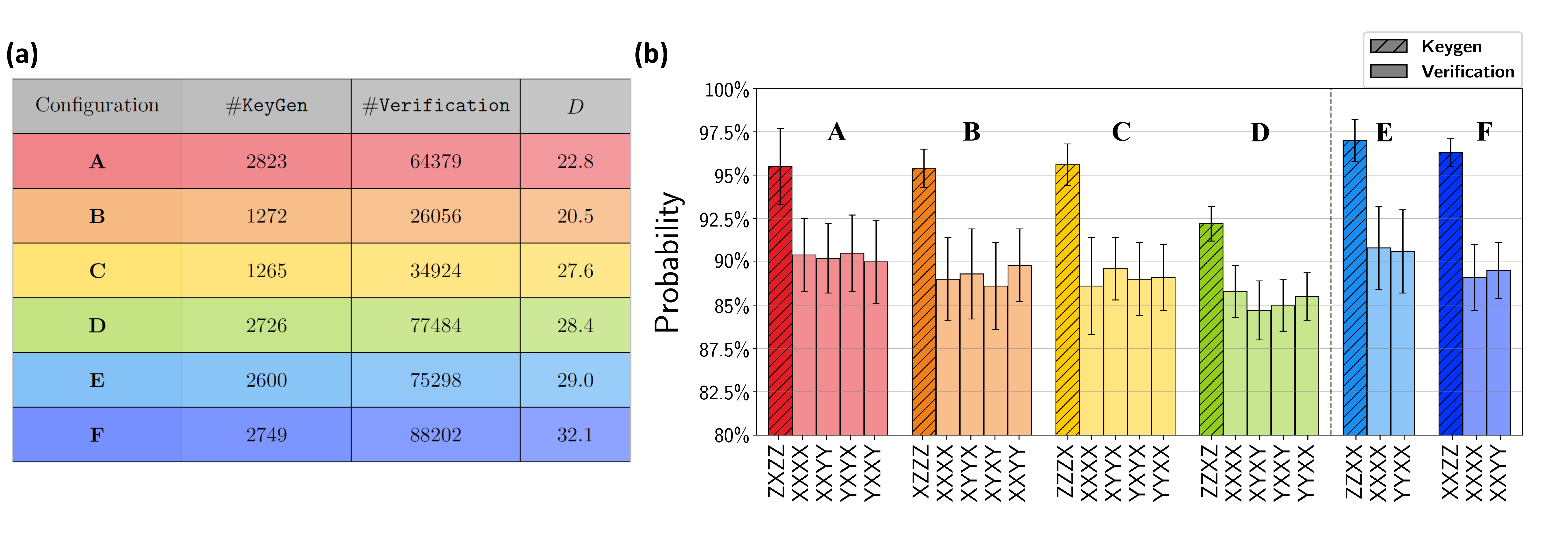}
         	\caption{\textbf{Experimental results.} \textbf{(a)} Total number of \texttt{KeyGen} and \texttt{Verification} rounds with corresponding security parameter $D$.
						\textbf{(b)} Success probability for key generation verification rounds for the different configurations. 
						We calculate the probabilities per run; the error bars show the standard deviation over the number of runs.}
 	\label{figure4}
\end{figure*}

For each configuration we run the protocol between 150 and 300 times, each run corresponding to a specific measurement setting. For each run, we integrate over $10\,$min of measurement time; each four-photon event is considered a round; this leads to the total number of \texttt{KeyGen} and \texttt{Verification} rounds given in Fig.~\ref{figure4} (a). 

We estimate the probability of a correct \texttt{KeyGen} round and a passed \texttt{Verification} for each run. We then calculate the averaged probabilities over all runs (see Fig.~\ref{figure4} (b)) for all measurement operators and all implemented three-partite (\textbf{A}, \textbf{B}, \textbf{C} and \textbf{D}) and bipartite (\textbf{E} and \textbf{F}) network configurations.
The success probability averaged over all the \texttt{KeyGen} runs is $95.3$ with a standard deviation of $\pm 1.3\%$, whereas for the \texttt{Verification} rounds the average success rate is $89.3(\pm 2.1)\%$.

The main source of noise in our implementation originates from higher-order emissions from the parametric down-conversion sources, which is about $3\%$ of all four-fold coincidences at a pump power of 180\,mW for each source. 
Furthermore, the generated Bell states show visibilities of about $\simeq0.97$ when both qubits are measured in either the $X$- or the $Z$-basis. At the polarising beam splitter, we achieve two-photon interference visibility of $0.823\pm 0.02$, mainly limited through higher-order emissions, residual distinguishability of the photons, and, in particular, imperfect mode overlap.

As displayed in Figure~\ref{figure4}(a), the effective security parameter $D$ ranges from about $20$ to $32$. 
If we just consider the number of \texttt{Verification} rounds vs. the number of \texttt{KeyGen} rounds, the probability $1/D$ of an adversary correctly guessing a round to be a \texttt{KeyGen} round  is on average about $4\%$ (see Appendix). In that case they can perform a $Z$ measurement instead of an $X$ measurement. Thus, they would effectively take part in the \texttt{KeyGen} round and could compromise the security of the key. Note that this is the probability per individual key bit, and all key bits are uncorrelated in this regard.

In our implementation, the probability of a successful \texttt{Verification} round is smaller than one. 
The worst-case scenario would be that all failed \texttt{Verification} rounds are accepted, but are in fact caused by a malicious adversary actively cheating. Then, the adversary can cheat during all these failed \texttt{Verification} rounds without being caught, thereby getting more `tries' to cheat during the \texttt{KeyGen} rounds. 
In that case, the adversary has $\eta(D-1)$ of these extra attempts, where $\eta$ the \textit{failure rate} of the \texttt{Verification} rounds. The average probability of the adversary correctly guessing the \texttt{KeyGen} round becomes then $\tfrac{1 + \eta(D-1)}{D}$. In our experiment, the adversary has a probability of $\sim 14$\% of correctly guessing each \texttt{KeyGen} round averaged over all different configurations (see Appendix for individual values).

However, in reality, the fidelity of the \GHZ{4} state is non-perfect, leading also to failed \texttt{Verification} rounds.
We can estimate the expected failure rate due to noise based on the (non-unit) fidelity of the distributed \GHZ{4} state. 
We look for a lower bound on the expected failure rate so that the adversary has the maximum number of cheating attempts based on our fidelity.
Using the trace distance and relating it to the fidelity, the lower bound on the expected failure rate can be estimated to be $r_{f} \geq 1 - \sqrt{F}$. We then replace $\eta$ by $\eta' = \eta - r_{f}$, or $0$ if this is negative.
With this correction the probability of the adversary guessing the \texttt{KeyGen} round without being caught reduces to $\sim7\%$ in our experiment (see Appendix for more information).

In summary, our analysis shows that our \texttt{KeyGen} error rates are on average $< 5\%$ and are thus correctable using standard approaches.
The probability of being correlated with the adversary can be bounded from above by $\lesssim 7\%$; additional correlations could be gained through error correction. Both contributions together are still expected to be within the limit that they can be reduced using privacy amplification~\cite{Grasselli2021}.
Note that our arguments are meant to give an estimate of the viability of the experimental implementation and the subsequent post-processing steps.
Performing these tasks in an anonymous fashion is a nontrivial task and further theoretical work is necessary to facilitate this.

Finally, we gather between 100 and 200 counts each run of 10 minutes integration time, which corresponds to $0.16$ - $0.33$\,bps. Taking into account that $1/30$ to $1/20$ are \texttt{KeyGen} rounds and the rest is \texttt{Verification} rounds, we get an effective keyrate between ~$0.006$ and $0.017$. The classical post-processing necessary to obtain a perfectly secure key is, as mentioned before, out of the scope of this article, but it will affect the effective key rate.


\section{\bf Discussion}
We have demonstrated how to anonymously and verifiably establish a shared key between several parties using multipartite quantum resources and exploiting the correlations of \GHZ{} states.
This is a significant step towards achieving secure and anonymous quantum communication, adding to the recent theoretical and experimental achievements in the field~\cite{Hahn2020, Unnikrishnan2019, Murta2020, Proietti2020, Huang2020}. 

For full-scale and real-life implementations of anonymous conference key agreement protocols, methods for error correction and privacy amplification need to be developed~\cite{Grasselli2019}
and, also, finite key effects to be considered~\cite{Proietti2020}.
Even though we make a first attempt to quantify the effect of experimental imperfections on the security of the protocol, further research is still needed.

In this context, the effect of losses on general conference key agreement has been studied recently~\cite{Singkanipa2021}; it would be interesting to investigate whether a similar approach can be deployed while preserving anonymity. 
Future steps will be the implementation of our protocol in larger-scale networks and active switching for closing loopholes~\cite{Huang2018}.
Finally, it will be interesting to see whether anonymity can be maintained with multipartite entangled resources other than \GHZ{} states.


\section{\bf Acknowledgements}
 We thank Jelena Mackeprang and Lukas R\"uckle for comments, and B\"ulent Demirel for setting up the early stages of the experiment.
A.P.~and J.d.J.~acknowledge support from the German Research Foundation (DFG, Emmy Noether grant No. 418294583) and F.H.~from the Studienstiftung des deutschen Volkes. C.T. and S.B~acknowledge support from the Carl Zeiss Foundation, the Centre for Integrated Quantum Science and Technology (IQ$^\text{ST}$), the German Research Foundation (DFG), the Federal Ministry of Education and Research (BMBF, project SiSiQ), and the Federal Ministry for Economic Affairs and Energy (BMWi, project PlanQK). 


\bibliographystyle{apsrev4-1}
%


\newpage
\onecolumngrid

\section*{Appendix}
\begin{figure*}[h]
\centering
	\includegraphics[width=0.75\textwidth]{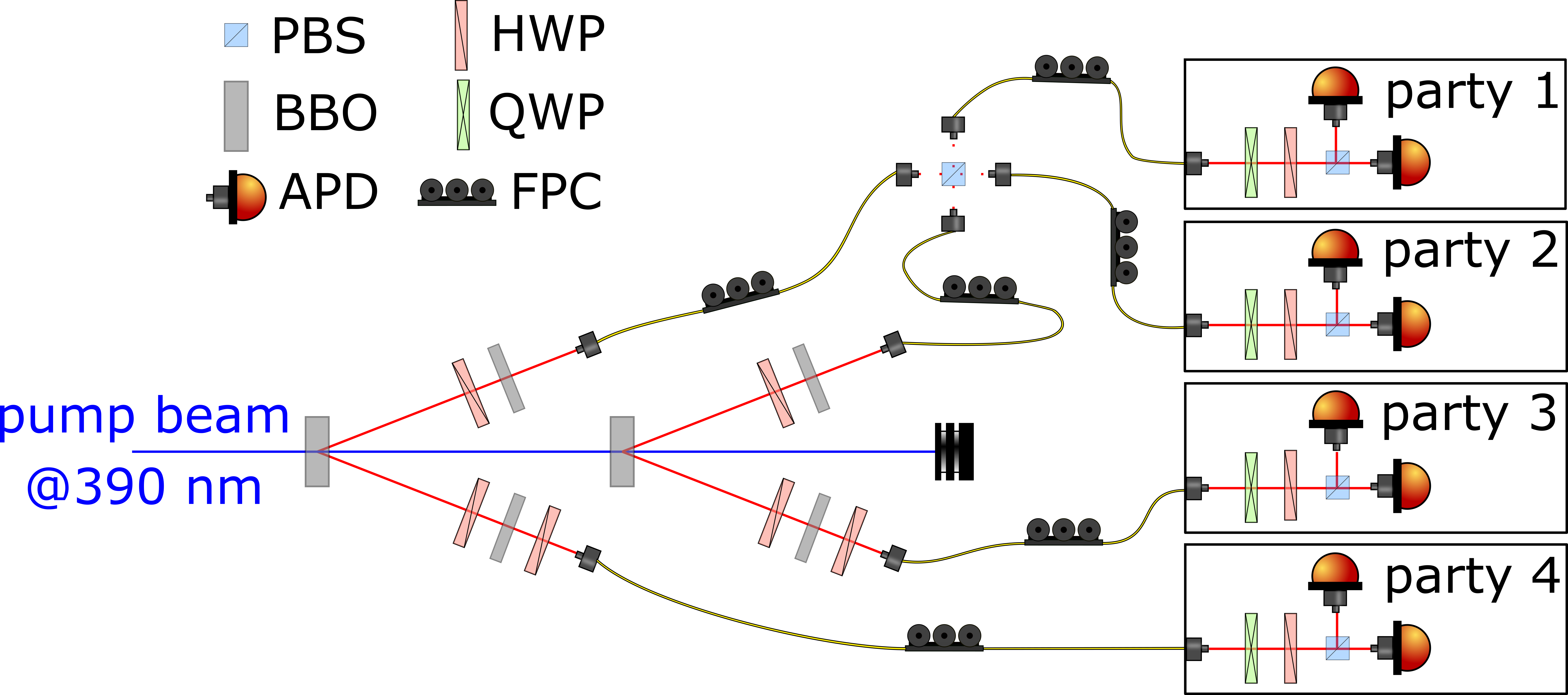}
	\caption{Experimental setup. Two $\beta$-barium borate (BBO) crystals are pumped with a fs-pulsed laser ($390\,$nm)
		to generate polarisation-entangled photon pairs via type-II spontaneous parametric down-conversion. After passing a half-wave plate and an additional BBO, the photons are prepared in state $\ket{\psi^-}=(\ket{H,V}-\ket{V,H})/\sqrt{2}$. Two photons, one from each source, are superimposed on a polarising beam splitter (PBS) to generate the four-photon state $\ket{\mathrm{GHZ}}=(\ket{H,H,H,H}+\ket{V,V,V,V})/\sqrt{2}$ after postselecting four-photon events. 
		The measurements as described in the main protocol are realised using quarter-wave (HWP) and half-wave plates (HWP), together with PBSs and avalanche photo detectors (APDs). FPC indicates fibre paddles for controlling the polarisation.}
	\label{figure5}
\end{figure*}

\begin{figure}
\centering
	\includegraphics[width=0.9\textwidth]{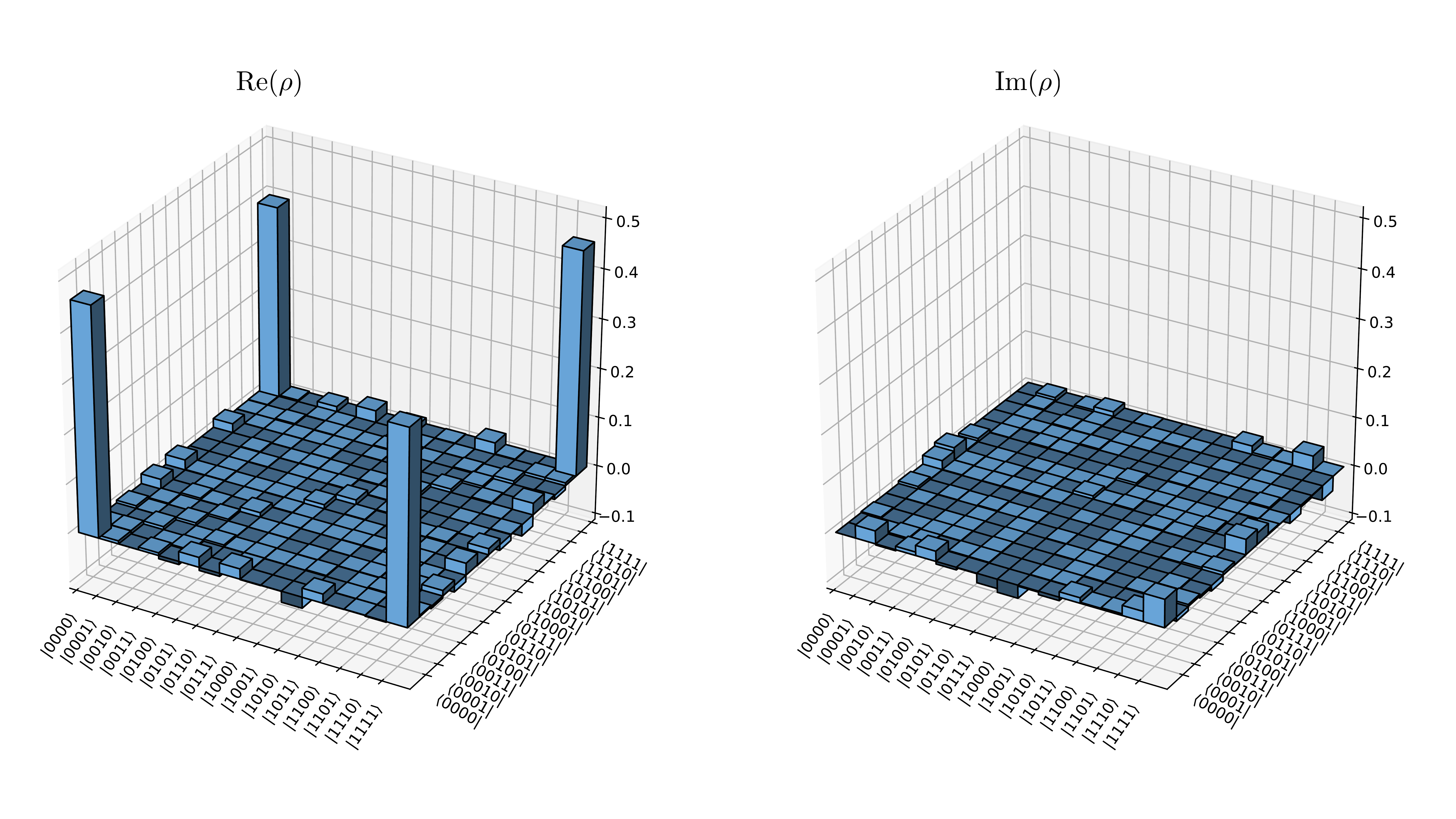}
	\caption{Reconstructed density matrix $\rho$ of the \GHZ{4} state used to implement the anonymous conference key agreement.}
	\label{figure6}
\end{figure}

\begin{figure}
\centering
	\includegraphics[width=0.9\textwidth]{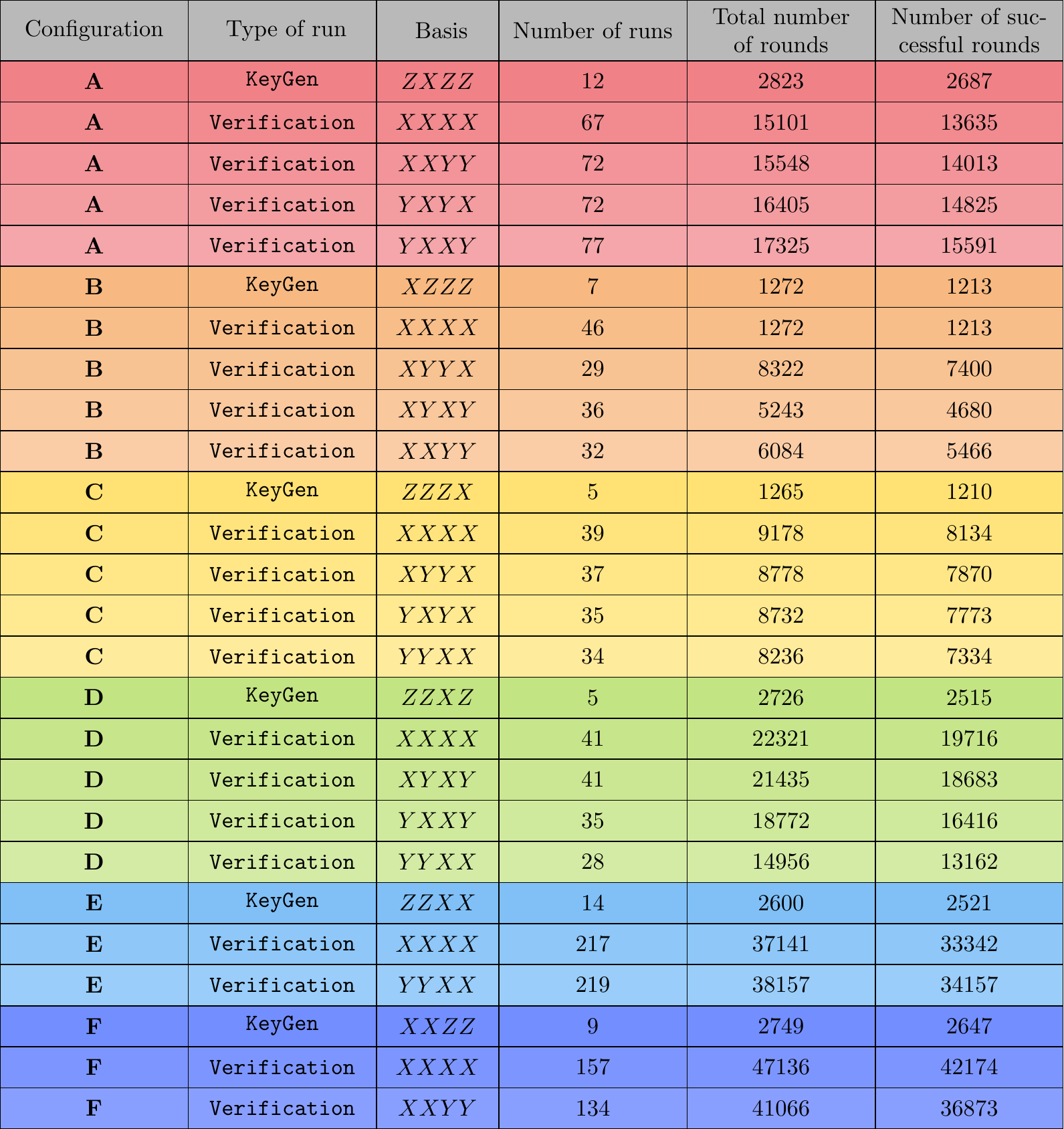}
	\caption{ List of all \texttt{KeyGen} and \texttt{Verification} runs for each configuration. Each configuration has a certain set of measurement bases for \texttt{KeyGen} and \texttt{Verification}. For each run, we integrate over 10~min of measurement time. The number of rounds refers to the total number of four-photon clicks in each basis, integrated over all runs. The number of successful rounds refers to the total number of four-photon clicks that correspond to a measurement result expected from the theory.}
	\label{figure7}
\end{figure}

\begin{figure}
\centering
	\includegraphics[width=0.9\textwidth]{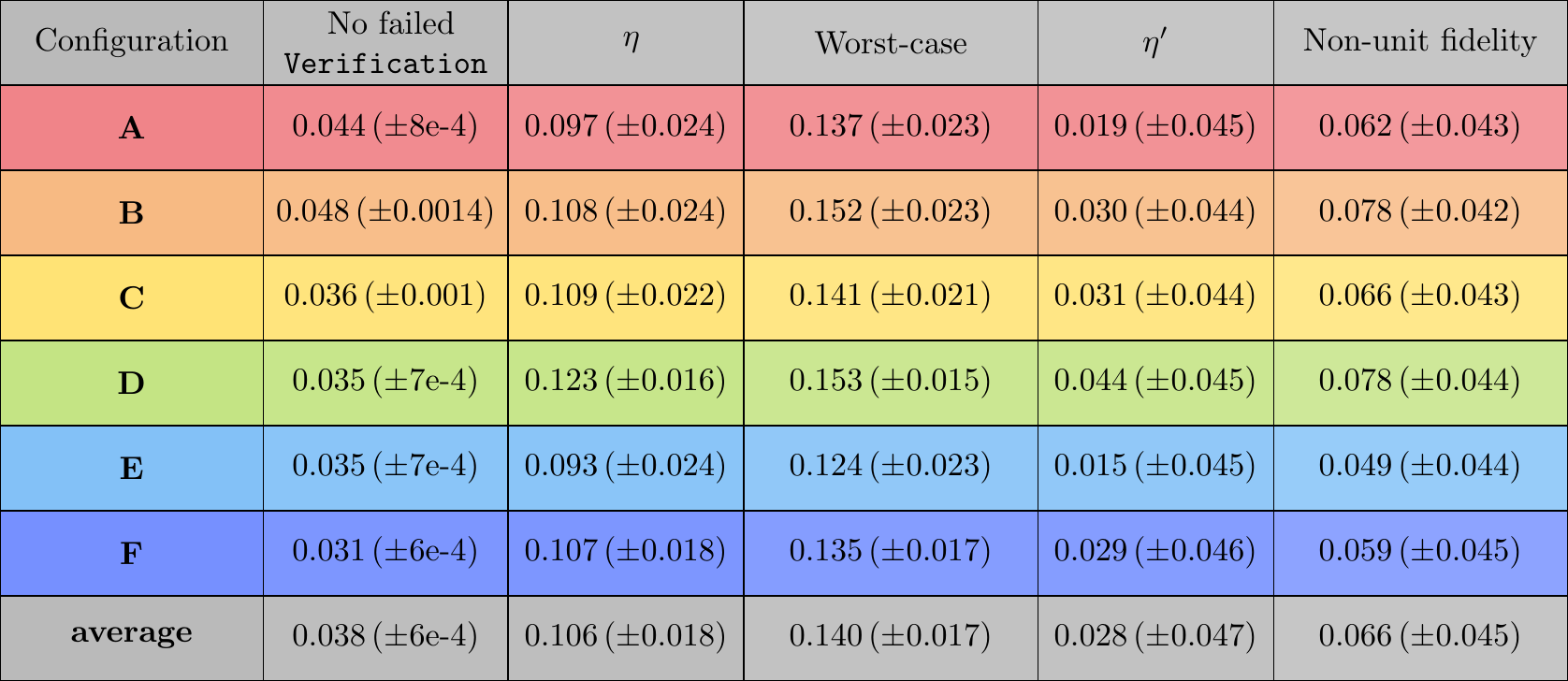}
	\caption{
		Probabilities for a potential adversary to correctly guess which round is the \texttt{KeyGen} round for each configuration and multiple scenarios. In the first scenario, corresponding to the second column ("No failed \texttt{Verification}"), we assume the guessing probability is $\frac{1}{D}$. This means we consider the number of \texttt{Verification} rounds vs. \texttt{KeyGen} rounds, meaning that effectively no failed
\texttt{Verification} is allowed.
		The second scenario, corresponding to the fourth column ("Worst case"), is a worst-case analysis, where all allowed \texttt{Verification} failures (from our experimental results) are assumed to come from a malicious, cheating, adversary - such that the guessing probability becomes $\frac{1+\eta(D-1)}{D}$, where $\eta$ is the failure rate of the \texttt{Verification} rounds. The third and final scenario takes a more realistic approach ("Non-unit fidelity"), where the non-unit fidelity of the shared \GHZ{4} state is taken into account. Here, $\eta$ is updated to $\eta'$ to correct for the fact that some of the \texttt{Verification} rounds will fail purely because of noise in the system---the updated guessing probability can be found in the sixth column.
These are heuristic analyses, but the guessing probability $p$ can be seen as (influencing) the information that the adversary possesses about the final raw key. These correlations with the adversary have to be deleted in post-processing, giving up a fraction $\sim h(p)$ of the raw key, where $h(p)$ is the binary entropy of $p$.}
	\label{figure8}
\end{figure}

Here, we give more information on the experimental setup for the implementation of the anonymous conference key agreement protocol.
The experimental setup for the generation of the four-photon GHZ state is depicted and described in Fig.~\ref{figure5}. Fig.~\ref{figure6} depicts the reconstructed density matrix of the \GHZ{4} state generated in the experiment.
We also add more detailed information re the security parameters of our implementation.
Fig.~\ref{figure7} shows the number of runs and rounds we perform in our experiment. In Fig.~\ref{figure8} we list the values that an adversary correctly guesses a \texttt{KeyGen} round for each configuration, assuming an ideal state or a non-ideal state.

\vspace{1cm}

%

%

%

\twocolumngrid

\end{document}